\input psfig
\documentstyle[twocolumn,prl,floats,aps]{revtex}
\newcommand{\be}{\begin{equation}}
\newcommand{\ee}{\end{equation}}

\begin{document}
\draft

%
%
\twocolumn[\hsize\textwidth\columnwidth\hsize\csname @twocolumnfalse\endcsname
\title{ A Possible Phononic Mechanism for $d_{x^2 - y^2}$-Superconductivity
        in the Presence of Short-Range AF Correlations}
\author{ Alexander Nazarenko and Elbio Dagotto}
\address{Department of Physics and National High Magnetic Field Lab,
Florida State University, Tallahassee, FL 32306, USA}
\maketitle
\begin{abstract}
\noindent
We discuss the high temperature superconductors in a regime where the
antiferromagnetic (AF) correlation length is only a couple of lattice
spacings. In the model proposed here, these short-range
AF fluctuations play an essential role in the
dressing of the carriers, but the attraction needed for
superconductivity (SC) arises from
a transverse phonon oxygen mode with a finite buckling
angle as it appears in ${\rm YBa_2Cu_3O_{7-\delta}}$.
A simple fermion-phonon model analog to the Holstein model is introduced
to account for this effect. We argue that the model has a
$d_{x^2 - y^2}$-wave superconducting groundstate.
The critical temperature ($T_c$) and the O-isotope
effect coefficient ($\alpha_O$) vs hole density ($x$) are in qualitative
agreement with experiments for the cuprates. The
minimum (maximum) of $\alpha_O$ ($T_c$)
at optimal doping is caused by a large peak in the density of states of
holes dressed by AF fluctuations, as discussed in previous van Hove scenarios.

\end{abstract}

]
%
%

  Since the discovery of high temperature superconductors, the origin
of their pairing mechanism has been controversial. Numerous studies
have shown that normal state properties deviate from a
conventional Fermi liquid, and as a possible explanation
several authors proposed the
AF correlations as responsible for such nonstandard behavior.
Theories based on AF pairing mechanisms
analyzed using diagrammatic and numerical techniques
predict $d_{x^2-y^2}$ SC.\cite{AF}
 Josephson junction experiments and angle-resolved
photoemission (ARPES) data are consistent with
such ${d_{x^2-y^2}}$ condensate.\cite{gap-symmetry,ARPES}

However, there are still some problems with this approach.
For example, the AF correlation
length, $\xi_{AF}$, in
the  normal state of the high-$T_c$ cuprates at optimal doping,
$x_{opt}$, may not be robust enough
to induce SC. NMR measurements in
${\rm YBa_2Cu_3O_{7-\delta}}$ suggest
$\xi_ {AF}/a \sim 2-3$\cite{NMR} (${ a}$ is the Cu-Cu lattice
spacing). Inelastic neutron scattering studies for the same compound
only show a broad peak at ${\bf q} =(\pi,\pi)$
in the dynamical spin structure factor.\cite{Neutron}
Weak ``shadow bands'' in ${\rm Bi2212}$ ARPES data  at room temperature
have been interpreted as produced by short-range AF correlations (although
superlattice effects have not been ruled out).\cite{Shadow}
Numerical studies\cite{haas} have shown that these weak AF-induced bands
are quantitatively reproduced by the doped t-J model in a
regime where $\xi_{AF}/a \sim 2-3$, in agreement with NMR.
While many
numerical studies suggest  that this apparently small
AF-correlation can nevertheless substantially affect the quasiparticle (q.p.)
dispersion (one-particle Green's function),\cite{flat}
it has not been shown that it can also lead to
 pairing (two-particle Green's
function, with hole Coulombic repulsion included)
as it may occur in the regime $\xi_{AF}/a \gg 1$.\cite{AF}

A second  problem for electronic mechanisms in general
is the O-isotope
effect observed in the cuprates. While experiments have shown that
the coefficient $\alpha_O$ is very small at ${ x_{opt}}$, it increases
in the underdoped and overdoped regimes
reaching values comparable to  the
BCS limit ${\rm \alpha_{BCS} = 0.5}$.\cite{franck}
It may be argued
that this effect is caused by spurious
changes in $x$ after the replacement
${\rm O^{16} \rightarrow O^{18} }$, but until a realistic calculation
proves it, the
experimental data cannot be simply neglected.

In this  paper we $assume$ that
the normal state $\xi_{AF}$ at $x_{opt}$ is not large enough
to produce magnon mediated pairing, and thus we discuss possible
alternative ideas that may explain SC in the cuprates.
We argue that a small $\xi_{AF}$ can still be important for the
normal state hole dispersion
since carriers are much affected by the surrounding
$local$ spin environment. Short AF correlations can modify the
hole dispersion reducing the bandwidth and producing anomalous ``flat-bands''
as observed in the t-J and Hubbard
models,\cite{flat} and in ARPES data.\cite{ARPES}
The flat-bands
induce a robust $peak$ in the density of states (DOS) at the top of
the valence band that boost
$T_c$ and produce an ``optimal doping'' when the chemical potential
$\mu$ reaches the peak, once a source of hole attraction exists.
This combination of the AF and van Hove scenarios\cite{afvh}
is similar in spirit to previously
discussed van Hove theories.\cite{isotope-effect} The key difference is
the origin of the DOS large peak  which in Ref.\cite{afvh}
was  attributed  to AF correlations.
In the present paper, pairing is caused by a
$phononic$-induced attraction supplemented by a hole
dispersion modulated by short distance AF correlations that tend to
prevent double occupancy and favor intrasublattice hopping.
Thus concepts of both phononic and electronic theories are here mixed
in a single scenario.

However, an immediate problem with this idea is the symmetry of the
SC condensate. Evidence is accumulating
in favor of  ${ d_{x^2 - y^2}}$ SC
which is natural in electronic AF-theories, but seems unnatural in
conventional phononic theories.
For example, the Holstein model couples electrons to
on-site oscillators\cite{holstein} leading to
a uniform s-wave condensate.
To obtain d-wave phononic SC,
we introduce a modification of the Holstein model with oscillators
located at the oxygens of a two dimensional (2D) square lattice
rather than at the copper sites
(Fig.1a,b). While this modification seems ``ad-hoc'', physical
realizations involving the buckling mode of oxygen in the cuprate
exist, as shown below. We further argue that the
effective hole-hole interaction produced by this model
favors d-wave SC at low carrier concentration
once the AF induced hole dispersion is used.
\begin{figure}[htbp]
\centerline{\psfig{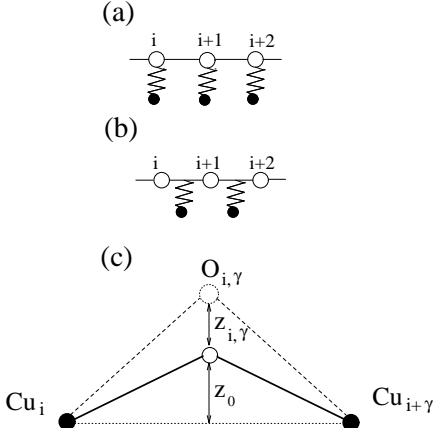}}
\caption{
(a) Graphical representation of the standard
Holstein model with an oscillator attached to each lattice site {\bf i};
(b) In the new model Eq.(1) the oscillators are attached to the lattice
Cu-Cu links (oxygen positions);
(c) Oxygen buckling mode studied in this paper. $z_0$ is the oxygen
equilibrium position,
and $z_{i,\gamma}$ is the displacement from equilibrium in the $\gamma$
direction.
}
\end{figure}


The modified Holstein model proposed here is
$$
{
H = \sum_{{\bf k}\sigma}[\epsilon_{AF}({\bf k})-\mu]
{\bar c}^{\dag}_{{\bf k}\sigma} {\bar c}_{{\bf k}\sigma} +
\sum_{{\bf i},\gamma=\hat{{\bf x}},\hat{{\bf
y}}}(\frac{p_{{\bf i}\gamma}^2}{2M}
+ \frac{M}{2} \omega^{2}z_{{\bf i},\gamma}^2)
}$$
$${
+ g \sum_{{\bf i},\gamma=\hat{{\bf x}},\hat{{\bf y}}}
(z_{{\bf i},\gamma}+z_{{\bf i},-\gamma})(\hat{n}_{{\bf
i}\uparrow}+\hat{n}_{{\bf i}\downarrow}), }
\eqno(1)$$
\noindent where $\epsilon_{AF}({\bf k})$ is the hole dispersion obtained
$after$ the influence of AF correlations has been taken into account
(this is important since the resulting q.p.'s are $weakly$ interacting and
thus the peak in the DOS is not removed by correlations
or disorder\cite{afvh}).
${\bar c}({\bar c}^{\dag})$ is a fermion destruction (creation)
operator located at Cu with standard fermionic anticommutation
relations (these fermions represent ``holes'' in the cuprates),
$\hat{n}_{{\bf i}\sigma}={\bar c}^{\dag}_{{\bf i}\sigma}
{\bar c}_{{\bf i}\sigma}$ is the
number operator, ${\bf i}$ labels sites of a 2D square lattice,
$\gamma={\bf {\hat x},{\hat y}}$  are unit vectors along the axis-directions,
and $z_{{\bf i},\gamma}$ and $p_{{\bf i},\gamma}$ are
the coordinate and momentum of the oscillators of mass $M$ and
frequency $\omega$
at half-distance between adjacent sites to mimic oxygens.
As hole dispersion we  use
$\epsilon_{AF}({\bf k})/eV = 0.165\,cosk_x\,cosk_y + 0.0435\,
(cos2k_x + cos2k_y)$, which
corresponds to the hole dispersion in an AF background at
half-filling,\cite{flat} but it should
be approximately
valid also at finite $x$ as long as
$\xi_{AF}$ is not negligible. Holes move within the same sublattice. The
spin index in Eq.(1) does not play an important role and working with
spinless fermions leads to similar results.
Using Eq.(1) we will study
small fermionic (hole) densities to mimic the
physics of the cuprates.


What cuprate
phonons can lead to Eq.(1)? The in-plane breathing
mode where O oscillates along the Cu-Cu link is not useful since it
produces an effective hole n.n. $repulsion$. In principle
O-phonons transverse to the Cu-Cu link have
a coupling quadratic in the displacement that is negligible.
However, several authors\cite{annett,normand}
noticed that the $buckling$ of the
Cu-O-Cu link leads to a linear electron-phonon coupling
in ${\rm Y Ba_2 Cu_3 O_{7-\delta} }$.
In this tilting mode (Fig.1c) the
O-atom oscillates in the {\bf \^{z}}-direction about an
equilibrium position $z_0$ with a buckling angle $\beta\,=\,5^{\circ}$ and a
frequency that here we take as
${ \hbar \omega_{buck}=12\,\,meV}$.\cite{cardona,annett}
While buckling effects are not
present in all cuprates, nevertheless it serves our purpose of
identifying at least one phononic mode that in combination with strong
correlations can lead to d-wave SC. Note that in this exploratory study
we are also neglecting the
coupling to other phonons modes of similar energy that may lead to repulsive
interactions.

Let us discuss recent
work related to the ideas described here.
Song and Annett\cite{annett} studied
d-wave phononic SC using the in-plane O-breathing mode. However,
further analysis showed that this mode does not lead
to d-wave SC.\cite{annett}
They also studied the O-buckling mode with a tight binding dispersion
finding d-wave SC, but
concluded that it would not
produce a large enough $T_c$.\cite{comm10}
The key difference with our approach is that we here use an AF-induced
DOS with a large peak which boots $T_c$ to high values. In other related work,
Yonemitsu et
al.\cite{yonemitsu} coupled the apical phonon modes to the t-J model
forming a SC polaron pair condensate. AF
correlations affect the hole propagation as in our approach, but
they found ``nodeless'' d-wave or p-wave pairing, contrary to our
$d_{x^2-y^2}$ result. Finally,
using a mean-field approximation for the t-J model with phonons,
Normand et al.\cite{normand} studied anomalies at $T_c$ associated to the
YBCO-buckling mode. $\alpha_O$ was reported
in agreement with experiment but the calculation was done
only at ${x_{opt} }$.

Returning to the main idea, let us derive the buckling-mode induced
hole-phonon coupling. Consider the Coulomb
energy of carriers at the Cu-ions
in the presence of the n.n. O-ions,\cite{Semicond}
$$
{
H_{Coulomb} = {{ e e^{\ast}}\over{\epsilon}}
\sum_{{\bf i},\sigma,\gamma}
{\bar c}^{\dag}_{{\bf i}\sigma} {\bar c}_{{\bf i}\sigma} (
{{1}\over{ | {\bf R}_{\bf i}-{\bf r}_{{\bf i},\gamma} | }} +
{{1}\over{ | {\bf R}_{\bf i}-{\bf r}_{{\bf i},-\gamma}| }} ),}
\eqno(2)
$$
where $e$ is the electron charge, $e^{\ast} = -2e$
is the O-ion charge, $\epsilon$ is the
dielectric constant, ${\bf R}_{\bf i}$ denotes the Cu-positions
which are assumed non-fluctuating, and ${\bf r}_{{\bf i},\gamma}$
denotes the vibrating O-positions (interactions at distances larger
than $a$ are assumed
negligible due to screening effects).
In Eq.(2) we define ${\bf r}_{{\bf i},-\gamma} =
{\bf r}_{{\bf i - \gamma},\gamma}$.
The Cu-O distance can
be expanded in the small O-ion displacement in
the $z$-direction, $z_{{\bf i},\gamma}$,
as, ${
\mid {\bf R}_{\bf i}-{\bf r}_{{\bf i},\gamma} \mid=
\frac{a}{2}(1+\frac{2z_0^2}{a^2})+
\frac{2z_0}{a}z_{{\bf i},\gamma}\,+\, \ldots, }$.
The relation ${ a \gg z_0 \gg z_{{\bf i},\gamma} }$ is assumed.
The hole-phonon interaction becomes
$$
{
H_{h-ph} = -
\frac{8ee^{\ast}z_0}{\epsilon a^3} \sum_{{\bf i}, \sigma,\gamma }
 {\bar c}^{\dag}_{{\bf i}\sigma}
{\bar c}_{{\bf i}\sigma}(z_{{\bf i},\gamma}+z_{{\bf i},-\gamma}).}
\eqno(3)
$$
In ${\bf k}$-space we arrive to
the fermion-boson Hamiltonian,
$$
{
H_{h-ph} = \frac{1}{\sqrt N} \sum_{{\bf kq}\gamma\sigma}g_
{{\bf q},\gamma}
 {\bar c}^{\dag}_{{\bf k}\sigma} {\bar c}_{{\bf k}-{\bf q}
\sigma}(b_{{\bf q},\gamma}+b^{\dag}_{-{\bf q},\gamma}), }
\eqno(4)
$$
where $N$ is the number of Cu-sites, ${ b_{{\bf q},\gamma}
(b^{\dag}_{{\bf q},\gamma}) }$ is the destruction (creation) phonon
operator with momentum ${\bf q}$,
and the electron-phonon coupling has the form
$$
{ g_
{{\bf q},\gamma}=-(\frac{16ee^{\ast}}{\epsilon a^2})
(\frac{z_0}{a})\sqrt \frac{\hbar}{2M \omega_{buck}}
\cos (\frac{q_{\gamma}}{2}), }
\eqno(5)
$$
where $M$ is the the O-mass.\cite{foot2}
The strength of the coupling is estimated as
$ g_{{\bf q}={\bf 0},\gamma} \approx (1-2)\cdot10^{-2}$ eV
where we used $\epsilon\,\approx\,10-20$\cite{epsilon}, $a\,\approx\,3.8\AA$,
$z_0\,\approx\,0.17\AA$\cite{cardona}, and
$M=16\,a.u.$. This is in
agreement with Ref.\cite{annett}, and it
is at least one order of magnitude $less$ than the typical
electron-phonon coupling strength
in normal metals, which is natural due to
the reduction caused by the geometric factor $z_0/a$.

Although a detailed study
of Eq.(4) would require the Eliashberg equations, here
we only analyze
the nonretarded version of the effective
hole-hole phonon-mediated interaction. This simplification
should not change the symmetry of the
SC condensate and other qualitative features
discussed below. Standard manipulations lead to the interaction
(natural units):
$$
{ H^{h-h}_{int}=
   \frac{1}{2N} \sum_{{\bf kpq}\sigma\sigma^{\prime}}V^{eff}_
{{\bf k},{\bf p}}
 {\bar c}^{\dag}_{{\bf p}\sigma}
 {\bar c}^{\dag}_{-{\bf p}+{\bf q}\sigma^{\prime}} {\bar c}_{-{\bf k}+{\bf q}
\sigma^{\prime}} {\bar c}_{{\bf k}\sigma}\,\,\,\,\,, }
\eqno(6)
$$
where $V^{eff}_{{\bf k},{\bf p}}$ is given by
$$
{
V^{eff}_
{{\bf k},{\bf p}}=-2\sum_{\gamma }
\frac{g^2_{{\bf p}-{\bf k},\gamma}
\omega_{buck}}{\omega_{buck}^2-
[\epsilon_{AF}({\bf p})-\epsilon_{AF}({\bf k}) ]^2}\,\,\,\,\,. }
\eqno(7)$$
Since we are considering a dilute gas of holes and a short-range
potential Eq.(7), this problem can be studied with the gap equation:
${ \Delta_{\bf p}=-\sum_{\bf k}
({V^{eff}_{{\bf k},{\bf p}} \Delta_{\bf k}}/{2E_{\bf k}})
\tanh \frac{E_{\bf k}}{2T}, }$
where $E_{\bf k}=\sqrt {[\epsilon_{AF}({\bf k})-\mu]^2+\Delta^2_{\bf
k}}$, and $\Delta_{\bf k}$ is the SC gap.
The  numerical solution of this
equation produces SC in the ${ d_{x^2-y^2}}$
channel. To obtain $T_c$ vs $x$
the linearized gap equation was solved.\cite{comm7}
Naively, the small hole-phonon coupling induced by the buckling mode
should produce a small ${\rm T_c}$. However, the
large peak in the hole DOS caused by flat bands
can boost $T_c$.\cite{afvh,isotope-effect}
The results are shown in  Figs.2 and 3.
$T_c$ is maximized when $\mu$ reaches the maximum in the
DOS producing an ``optimal doping''.\cite{isotope-effect}
For the couplings and the approximations described in this paper, which
are standard, $T_c$ reaches 30K which can be made higher by tuning the
q.p. dispersion to increase the boosting DOS.
Thus, the experimental symmetry of the SC phase,
and the $x$-dependence of $T_c$  can be qualitatively
reproduced by the small fermion-phonon coupling induced by the
buckling mode of YBCO,
if $\epsilon_{AF}({\bf k})$ is used as dispersion.
\begin{figure}[htbp]
\centerline{\psfig{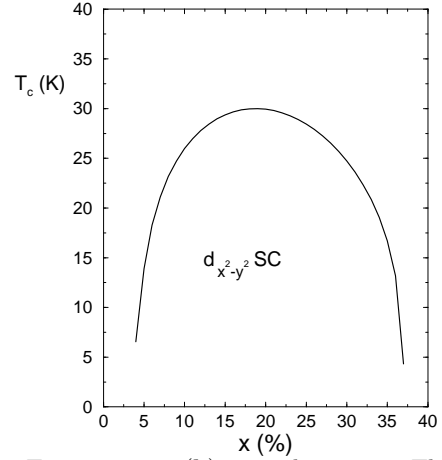}}
\caption{
${ T_c}$ vs ${x}$ using $\epsilon({\bf k})_{AF}$ as dispersion.
The coupling
$ g_{{\bf q}={\bf 0},\gamma}$ is set to 0.018 eV. The SC
state is d-wave.}
\end{figure}

Finding d-wave SC from
Eqs.(6,7) is natural. To visualize this effect, assume interacting
on-shell holes. Then,
transforming Eq.(6) into real space we get
$$
{
H^{h-h}_{int}=
  - \sum_{\bf i}
\frac{g^2_{{\bf 0},\gamma}}{\omega_{buck}}
\hat{n}^2_{\bf i}
  - \frac{1}{2} \sum_{\langle {\bf i j} \rangle}
\frac{g^2_{{\bf 0},\gamma}}{\omega_{buck}}
\hat{n}_{\bf i}\hat{n}_{\bf j}\,\,\,\,\,, }
\eqno(8)
$$
where $\langle {\bf i j} \rangle$ denote n.n. sites.
The first term in Eq.(8) is an
on-site attraction which is suppresed trivially by the hard-core
properties of the holes in the t-J model from which the
dispersion is derived.\cite{flat}
The second term provides the n.n. attraction that leads to an
interaction of the form corresponding to the
well-known ``t-U-V'' (${\rm U>0,V<0}$) model that has the tendency to form
d-wave condensates.\cite{tuv} However, there are several crucial differences
between our model and the t-U-V model. More remarkable is the fact that
the AF dispersion
used here allows the formation of a d-wave condensate at $low$ particle
density, while the t-U-V model (where a $cosk_x + cosk_y$ dispersion is
used) has s-wave SC in this regime\cite{tuv}. Optical conductivity
measurements in underdoped cuprates clearly show that
the number of carriers grows like the number of holes, and thus we
should study a dilute gas of quasiparticles as carried out in this paper.
Our ideas go beyond previous studies of d-wave SC that have used the
t-U-V model in spite of its shortcomings like having a $T_c$ maximum at
half-filling and a strong competition with phase separation.\cite{tuv}
Combining the potential Eq.(7) with the AF
dispersion\cite{flat} is the proper way to mimic the phenomenology of the
cuprates. In addition, the phonons in Eq.(1) produce an isotope effect
which does
not exist in the electronic t-U-V model used before in the literature.
Note also that
the instantaneous Coulombic repulsion at distance $a$
would tend to suppress an effective attraction of electronic origin,
like
magnon mediated interactions, since it does not lead to a substantial
retardation effect. On the other hand, the retardation intrinsic to
phononic mechanisms avoids such a  problem in the present approach.

\begin{figure}[htbp]
\centerline{\psfig{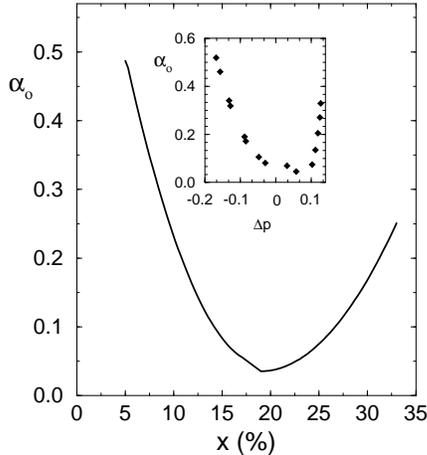}}
\caption{
$\alpha_O$
vs $x$ using $\epsilon({\bf k})_{AF}$, and $ g_{{\bf q}={\bf 0},\gamma}
= 0.018$ eV.
The inset shows experimental results for $\alpha_O$ vs $\Delta p$,
which is the concentration of
mobile holes per CuO plane relative to the optimal value corresponding to
${\rm Y Ba_2 Cu_3 O_7}$ when doping with Pr, Ca, Zn and/or Co reduces
$T_c$ (taken from J. P. Franck et al.,
Phys. Rev. {\bf B 44}, 5318 (1991);
J. P. Franck et al., in {\it Lattice Effects in High-$T_c$
Superconductors}, World Scientific (1992), p. 148; and references therein).}
\end{figure}

We also remark that
$\alpha_O \, \approx$$- (\Delta T_c / T_c) (M/\Delta M)$
vs $x$ has the proper shape compared to
experiments (Fig.3). The isotope coefficient
can be as low as 0.05 at $x_{opt}$. The minimum in $\alpha_O$
is caused by the
van Hove singularity in the dispersion, as remarked in
previous papers.\cite{isotope-effect}
Away from $x_{opt}$, $\alpha_O$ recovers the
value close to 0.5 as in standard phononic systems.
The behavior of $\alpha_O$ is regulated equally by $\Delta T_c$ and
$T_c$ i.e. $\Delta T_c$ vs $x$ is not constant in our calculation, but
minimized at $x_{opt}$.

Summarizing, here we have proposed
a model where phononic
pairing occurs between holes that are
strongly dressed by AF fluctuations. The model
may be applicable to the cuprates if the normal state correlation
$\xi_{AF}$ at $x_{opt}$ is proven not strong enough
to produce pairing (currently under much discussion).
Within the gap equation formalism, we found $T_c \sim 30K$ when
the buckling mode of YBCO is considered.\cite{nume} Although the hole-phonon
coupling is much
smaller than in normal metals, the large hole DOS boots $T_c$ to
realistic values. The same effect leads to an O-isotope coefficient
that is small at $x_{opt}$ but becomes close to 0.5 in the
underdoped and overdoped regimes. The symmetry of
the condensate produced by the buckling mode in $combination$ with the AF
induced hole dispersion is $d_{x^2 - y^2}$ even in the low density of
carriers regime.
These ideas may provide a tentative unified explanation for several puzzling
experimental features observed in the cuprates, especially the presence
of an abnormal Fermi liquid at $T> T_c$ coexisting with a nonzero
isotope effect.


We thank J.  F. Annett, J. P. Franck and M. Norman for useful correspondence.
Support by the Office of Naval Research under
grant ONR N00014-93-0495 and by the NHMFL is acknowledged.



\begin{references}
\bibitem{foot2}
 Note that the coupling $g_
{{\bf q}, \gamma}$ Eq.(5) does not depend on
${\bf k}$.
A ${\bf k}$-dependence appears if, e.g.,
the influence of the buckling mode on
the amplitude for n.n. hopping is considered using
$t(r)=t(\frac{a}{r})^\beta$, where $r$ is the Cu-O fluctuating distance
and $\beta \sim 6-10$ according  to some experiments.
\bibitem{comm7} $\alpha_O$ was calculated using an energy cutoff at
$\omega_{buck}$ around the Fermi surface, and a reduced effective
potential obtained from Eq.(7) by neglecting the momentum dependence in
the denominator.
\bibitem{nume} We encourage
numerical work on Eq.(1)
to obtain information beyond the gap equation.
\end{references}
\end{document}